# Artificial Intelligence-Driven Prognostic Classification of COVID-19 Using Chest X-rays: A Deep Learning Approach

Alfred Simbun [1] and Suresh Kumar [2] *

[1] Post Graduate Centre, Management and Science University Seksyen 13, 40100 Shah Alam, Selangor, Malaysia.

[2] Faculty of Health and Life Sciences, Management and Science University, Seksyen 13, 40100 Shah Alam, Selangor, Malaysia.

* **Corresponding email:** sureshkumar@msu.edu.my

**ABSTRACT**

**Background:** The COVID-19 pandemic has overwhelmed healthcare systems, emphasizing the need for AI-driven tools to assist in rapid and accurate patient prognosis. Chest X-ray imaging is a widely available diagnostic tool, but existing methods for prognosis classification lack scalability and efficiency. **Objective:** This study presents a high-accuracy deep learning model for classifying COVID-19 severity (Mild, Moderate, and Severe) using Chest X-ray images, developed on Microsoft Azure Custom Vision. **Methods:** Using a dataset of 1,103 confirmed COVID-19 X-ray images from AIforCOVID, we trained and validated a deep learning model leveraging Convolutional Neural Networks (CNNs). The model was evaluated on an unseen dataset to measure accuracy, precision, and recall. **Results:** Our model achieved an average accuracy of 97%, with specificity of 99%, sensitivity of 87%, and an F1-score of 93.11%. When classifying COVID-19 severity, the model achieved accuracies of 89.03% (Mild), 95.77% (Moderate), and 81.16% (Severe). These results demonstrate the model's potential for real-world clinical applications, aiding in faster decision-making and improved resource allocation. **Conclusion:** AI-driven prognosis classification using deep learning can significantly enhance COVID-19 patient management, enabling early intervention and efficient triaging. Our study provides a scalable, high-accuracy AI framework for integrating deep learning into routine clinical workflows. Future work should focus on expanding datasets, external validation, and regulatory compliance to facilitate clinical adoption.

## INTRODUCTION

COVID-19, also known as the novel coronavirus, is a highly infectious respiratory illness caused by the severe acute respiratory syndrome coronavirus 2 (SARS-CoV-2). The outbreak of COVID-19 was first reported in Wuhan, China in December 2019 and has since spread rapidly to become a global pandemic [1, 2]. Symptoms of COVID-19 can range from mild to severe and can include fever, cough, shortness of breath, fatigue, body aches, and loss of taste or smell [3, 4, 5]. In some cases, COVID-19 can lead to severe respiratory illness, including pneumonia, acute respiratory distress syndrome (ARDS), and death [6, 7]. Preventative measures to reduce the spread of COVID-19 include wearing masks, practicing good hand hygiene, maintaining physical distance from others, and getting vaccinated.

Artificial intelligence (AI) has emerged as a valuable tool in the fight against COVID-19, with applications spanning various areas. However, it is important to acknowledge and address the limitations and challenges associated with the use of AI in this context. One significant challenge lies in the accuracy and potential bias of AI-based diagnostic tools. These tools, designed to assist in the diagnosis of COVID-19 based on medical images, have been subject to scrutiny. Concerns have been raised about the reliability of publicly available datasets used to train these models [13]. Common mistakes, such as failing to establish bona fide sources of datasets, removing duplication from sources, and pre-processing data into non-DICOM formats, can lead to biases and compromised model performance [14, 15]. Ensuring the use of high-quality, diverse, and representative datasets is essential to mitigate these limitations. By addressing these challenges through rigorous data collection, vetting reliable sources, and robust data cleaning processes, the accuracy and performance of AI-based diagnostic tools can be improved.

In addition to diagnostic tools, AI has been used for contact tracing [16], predictive modeling, compliance monitoring [17], and social media surveillance in the context of COVID-19 [18]. However, there are limitations and challenges associated with these applications as well. Monitoring compliance with safety measures using AI-powered cameras and sensors raises ethical concerns related to privacy and surveillance. Care must be taken to ensure that the deployment of such technologies respects individual rights and privacy protections. Transparency, informed consent, and robust privacy safeguards should be integral to the design and implementation of monitoring systems. Furthermore, social media monitoring, while helpful in identifying and countering misinformation, also requires careful consideration of ethical guidelines to avoid overreach and suppression of free speech. Striking the right balance between identifying misinformation and avoiding unintended biases or censorship is crucial in maintaining a trustworthy and inclusive information ecosystem [19].

Moreover, the reliance on AI models requires large and diverse datasets to learn effectively. In the case of COVID-19 X-ray classification, the scarcity of labeled images due to limited availability of confirmed COVID-19 cases can lead to imbalanced datasets and biased models that struggle to generalize well to different populations or variations in X-ray imaging techniques. Additionally, the reliability of publicly available data sources used to build deep learning predictive models has been called into question [20]. These challenges highlight the need for establishing trusted sources for datasets and ensuring robust data cleaning practices to mitigate biases and enhance the performance and generalizability of AI models. As the COVID-19 pandemic evolves and new technologies emerge, the utilization of AI in the fight against the virus will continue to advance. By addressing the limitations and challenges discussed above, such as ensuring dataset reliability, mitigating biases, upholding privacy and ethical guidelines, we can harness the full potential of AI to effectively prevent the spread of COVID-19 while maintaining trust, fairness, and privacy.

**MATERIALS AND METHODS**

*i. Pre-processing*

To ensure the clinical and computational reliability of our deep learning predictive model, we obtained a comprehensive dataset consisting of 1,103 confirmed Covid-19 positive cases' x-ray images. These images were originally in DICOM format and were sourced from the AIforCOVID database [21]. The dataset incorporates data and clinical information from six Italian hospitals, specifically collected during the initial wave of the COVID-19 emergency in Italy, spanning from March to June 2020. The images and associated data were generated as part of routine clinical care for symptomatic patients who were hospitalized due to COVID-19. Importantly, the dataset underwent a retrospective review and was collected following the careful anonymization of patient information to ensure privacy and confidentiality. Ethical approval for the study was obtained from the Ethics Committee, with the Trial-ID being 1507 and the approval date being April 7th, 2020. Notably, the dataset includes a gender distribution of 382 male and 722 female records, and it is further categorized into Moderate Covid (90 patients), Mild Covid (447 patients), and Severe Covid (569 patients) confirmed cases. The age range of the patients varies across different age groups, spanning from 18 to 100 years old.

Since the aim of this study is to classify the prognosis of Covid-19 infection, a total of 576 x-ray images publicly classified as Normal healthy patients [22, 23], and another 576 x-ray images publicly classified as Bacterial Pneumonia [23], Viral Pneumonia [24, 25] and Lung Opacity [25] were downloaded from Kaggle.com and included in the analysis. Prior to finalizing the number of images for each category of Covid-19 confirmed cases together with Normal and Viral Pneumonia, Bacterial Pneumonia and Lung Opacity datasets, several analyses were performed to review the level of difference between individual classes. The images underwent a manual review process to determine their suitability as a training set for the model. The review involved considering the following criteria: *i*) whether the image was excessively dark or light, *ii*) whether the x-ray was rotated in relation to the anatomical position, and *iii*) whether the x-ray contained significant areas of black. When the detector detected insufficient x-rays beams, this will cause the x-ray image intensity becomes lighter (too white), while too much radiation has been administered for the image acquisition, the resulting image will be too dark (black). Both errors diminish the diagnostic utility of the x-ray (Figure 1).

**Figure 1:** *X-ray images of too light and too dark.*

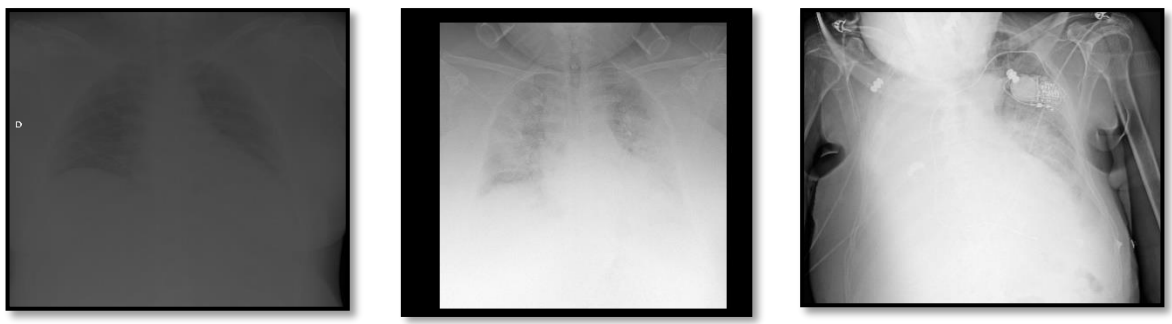

Images should be acquired in anatomic position, when possible, hence standardizing patient positioning is important for image interpretation, particularly in the field of medical imaging. By acquiring images in anatomic position, the appearance of the images becomes more predictable, allowing the image interpreter to develop a better understanding of what normal images look like. This can greatly assist in the detection of abnormalities, as the interpreter can easily differentiate between normal anatomy and any deviations from it (Figure 2). X-ray images that contain large black area were cropped manually to obtain areas that are showing the chest area (Figure 3). Subsequently, a total of 124 disqualified x-ray images were removed from the dataset leaving a total of 451 x-ray images belonging to severe cases, 372 x-ray images belonging to mild cases and 80 x-ray images belonging to the moderate cases. Each x-ray image was scalled and standardized into a 255 x 255 dimension allowing unbias image pattern learning during model development.

These x-ray images were further enhanced through an augmentation method, an image processing technique that is commonly used in computer vision and deep learning tasks to artificially increase the diversity and variability of a dataset by applying various transformations to the existing images. These transformations can include geometric operations, color manipulations, and other modifications. The purpose of image augmentation is to create new training examples that are slightly different from the original images, while still preserving the underlying features and labels. Image augmentation technique has helped to create 10 different random rotation augmentations with 50 degrees for each image (Figure 4) which gave 880 images for class moderate, 4,092 images for class mild, 4,961 images for class severe, 6,336 images for class normal, viral pneumonia, bacterial pneumonia, and lung opacity respectively.

**Figure 2:** *X-ray images of abnormal positioning*

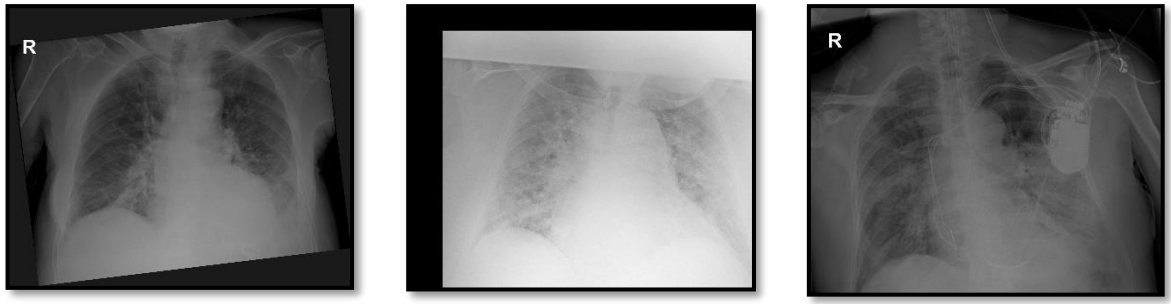

**Figure 3:** *X-ray images of with acceptable quality.*

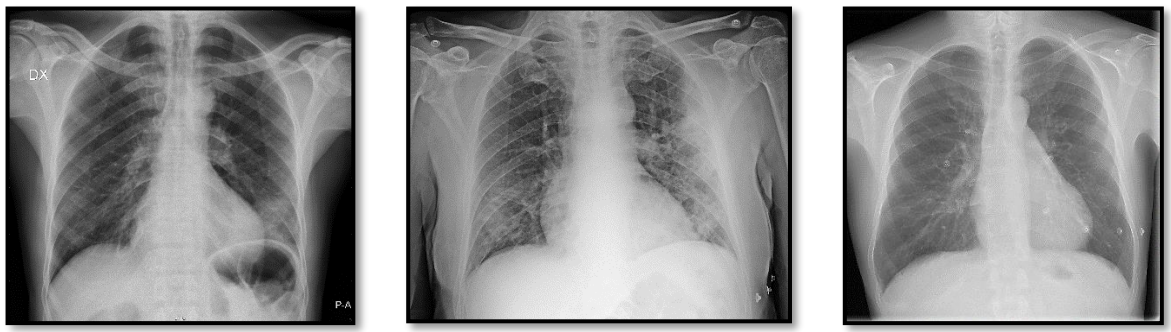

**Figure 4:** Examples of augmented *X-ray images of with acceptable quality.*

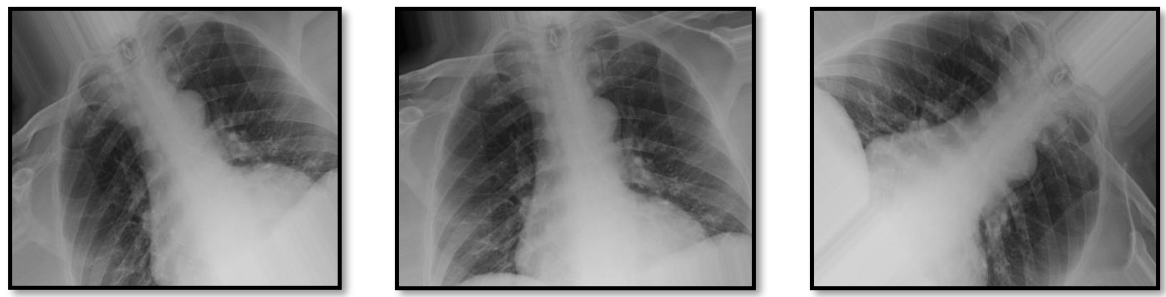

To achieve balanced datasets, the number of images in each class, except for the moderate class, was specifically adjusted to match the total quantity of x-ray images present in the moderate class, which was 880 images. This adjustment ensured that each class had a similar number of images for a more equitable distribution. We further divided these images into 70 percent training set (616 images) and 30 percent testing set (264 images). The x-ray images were divided into three distinct datasets, each containing a varying number of images (Table 1). Dataset Set 1 contains only the original number of x-ray images as imbalanced training set without augmentation, while both datasets Set 2 and 3 contains 70% balanced and augmented training set. 30% testing dataset were used to test out AI model Sets 1, 2 and 3 (Table 6). The goal was to determine which dataset yielded superior precision and recall outcomes for COVID-19 prognosis classifications based on image count and preprocessing profile and differences of training hours per iteration. This analysis potentially contributes to the development of a predictive model that can accurately classify and differentiate between different respiratory conditions, including COVID-19.

Table 1 shows information about three different datasets, each with a different number of classes and different properties related to data augmentation and balance. "Classes" refers to the number of different categories or labels in the dataset. "With augmented images" indicates whether the dataset has been augmented by adding modified or synthetic images to increase the diversity of the data. "Balanced" indicates whether the dataset has an equal number of examples for each class or label, which can be important for preventing bias during training. Set 1 has 2,055 images in total cases. It is not augmented and inbalanced. Set 2 has 616 examples for each class excluding Bacterial Pneumonia and Lung Opacity, has augmentation, and is balanced.

ii.  *Modelling using Custom Vision*

The accuracy of AI prediction relies on the detection of patterns and learning capability of this model. Theoretically, the more images the model has seen, prediction's precision and accuracy become higher. Hence, giving users a reliable source of confident. However since the reliability of publicly confirmed Covid-19 datasets are still questionable especially those from common sources, this study is compelled to use confirmed Covid-19 positive cases datasets from AIforCOVID on the basis of i) minimizing model's prediction bias towards unregulated public datasets, and ii) ensuring clinically reliable proposed model and its' prediction's accuracy. Microsoft Azure Custom Vision is used to build the high precision and recall image classification model using the same datasets (Table 1) with the following parameters: General (compact) [S1] as the chosen domain since compact domains are lightweight models that can be exported to iOS/Android and other external platforms for downstream analysis. Multiclass (single tag per image) is chosen since we only need to classify an image correctly into either one of the classes. Hence, there are three AI models were created using Microsoft Azure Custom Vision, each corresponding to individual dataset (Table 1) with several iterations and specific training hours (Table 2). The purpose of having these AI models is to aid in shedding insights into the sensitivity and accuracy of models created by Microsoft Azure Custom Vision. This helps to further understand the reliability of no-code computer vision tool that can accelerate prognosis classification based on iteration counts, number of classes and type of datasets used to generate these models.

iii.  *Internal and External Validations*

The evaluation of trained models often entails the utilization of essential performance metrics, including accuracy (AC), specificity (SP), sensitivity (SE), and precision (PR). These metrics play a crucial role in assessing the model's effectiveness in classification tasks. Accuracy (AC), calculated as (TP + TN) / (TP + TN + FP + FN), quantifies the proportion of correct predictions over the total number of instances. Specificity (SP), given by TN / (TN +

FP), measures the model's ability to accurately identify negative instances. Sensitivity (SE) or recall, computed as TP / (TP + FN), gauges the model's capability to correctly detect positive instances. Precision (PR), defined as TP / (TP + FP), examines the precision of positive predictions. The F1 score, a widely used metric, considers both precision and recall, providing a balanced assessment of the model's performance. It is computed as the harmonic mean of precision and recall using the formula 2 * (Precision * Recall) / (Precision + Recall). It is important to note that the calculation of true positives (TP), true negatives (TN), false positives (FP), and false negatives (FN) relies on a predetermined prediction threshold, typically 0.5, which distinguishes positive and negative predictions. While Microsoft Custom Vision offers evaluation metrics such as Precision, Recall, and Mean Average Precision for models based on seen or training datasets, it does not directly provide the Receiver Operating Characteristic (ROC) curve and Area Under the Curve (AUC) metrics within its interface. Consequently, assessing a model's performance on unseen datasets by considering TN, TP, FN, and FP enables a thorough examination of its generalization and predictive capabilities.

# RESULTS

**Table 1:** *Different datasets used to investigate image patterns and relationships between classes*

| Dataset | Classes | | | | | | | With augmented images | Balanced |
|---|---|---|---|---|---|---|---|---|---|
| | Severe Covid | Mild Covid | Moderate Covid | Normal | Viral Pneumonia | Bacterial Pneumonia | Lung Opacity | | |
| Set 1 | 451 | 372 | 80 | 576 | 576 | 0 | 0 | No | No |
| Set 2 | 616 | 616 | 616 | 616 | 616 | 0 | 0 | Yes | Yes |
| Set 3 | 616 | 616 | 616 | 616 | 616 | 616 | 616 | Yes | Yes |

Table 1 provides an overview of different datasets used for this study. Hypothetically, dataset Set 1 has a moderate number of images but is not fully balanced, potentially impacting the model's performance for specific Covid-19 classes. Set 2 is a balanced dataset but lacks information about the source and diversity of the additional images (excluding Bacterial Pneumonia and Lung Opacity). Set 3 is a large, balanced dataset but may require significant computational resources.

**Table 2:** *Different datasets used to investigate variance in performance metrics.*

| AI Model | Dataset | Iteration | Training Hour | Overall Precision | Overall Recall | Overall A.P. | F1 Score |
|---|---|---|---|---|---|---|---|
| 1 | Set 1 | 1 | 0 | 94.60% | 94.40% | 96.30% | 94.50% |
| | | 2 | 6 | 94.60% | 94.60% | 98.10% | 94.60% |
| 2 | Set 2 | 1 | 0 | 87.30% | 87.20% | 91.60% | 87.25% |
| | | 2 | 6 | 91.90% | 91.70% | 94.60% | 91.80% |
| 3 | Set 3 | 1 | 0 | 88.60% | 88.20% | 94.30% | 88.40% |
| | | 2 | 6 | 93.19% | 93.04% | 96.10% | 93.11% |

**Table 2** provides information about three different AI models, including the dataset used, iteration number, training hour, overall precision, overall recall, and overall A.P. (average precision) and F1 score. All three AI models underwent two iterations with iteration 1 being trained for 0 hours, and iteration 2 being trained for 6 hours. The table shows that the model's performance varies depending on the dataset used, with Set 2 achieving the highest overall precision, recall, and A.P. and F1 scores for both iterations. AI Model 1 was trained on Set 1 and has gone through two iterations of training for 0 and 6 hours each. The model's overall precision, recall, and A.P. and F1 score were calculated for each iteration. The highest precision and recall were achieved at the second iteration with 94.60% precision and recall, and the highest A.P. was achieved at the same iteration with 98.10%. The F1 score for AI model Set 1 is 94.60%. Similarly, AI Model 2 was trained on Set 2 with varying iterations and training hours, and achieved its highest overall precision, recall, and A.P. at the second iteration with an F1 score at 91.80%. For AI Model 3, the highest overall precision, recall, and A.P. were achieved at the second iteration of Set 3 with an F1 score at 93.11%. However, it is important to know that AI model Set 1 and Set 2 do not include Bacterial Pneumonia and Lung Opacity classes during model training whereas AI model Set 4 has seven classes which include Bacterial Pneumonia and Lung Opacity (Table 1). A longer training hour (beyond 6 hours) could not be performed since longer training hours on Microsoft Azure Custom Vision is considerably costly.

From the provided table, we can gain several insights on how the AI models performed on different datasets and iterations. The overall precision and recall for each AI model show how well it can predict positive cases (precision) and how well it can identify all positive cases (recall). We can see that Set 1 iteration 2 has the highest A.P. (98.10%), while Set 2 iteration 1 has the lowest A.P. (91.60%). The A.P. shows how well the AI models rank the positive cases. In general, we see that the A.P. increases as the training hours increase. The F1 score is a harmonic mean of precision and recall and provides a balanced measure of the two. Set 1 iteration 2 has the highest F1 score (94.60%) while Set 2 iteration 1 has the lowest F1 score (87.25%). However, it is also worth noting that the results may depend on the specific dataset and the performance metrics used. The AI models generally exhibit consistent performance across iterations within the same dataset. For example, dataset Set 1 shows consistent precision, recall, A.P., and F1 score between iterations 1 and 2. In most cases, increasing the training hours leads to improved performance metrics. This indicates that more training allows the models to better learn and generalize patterns from the data. Each dataset has its own characteristics, and this can affect the performance of the AI models. For instance, dataset Set 2 initially performs relatively lower compared to the other datasets in terms of precision, recall, A.P., and F1 score. There is variation in the performance metrics across the different datasets and iterations. This suggests that the choice of dataset and training iteration can significantly impact the model's performance. The performance metrics at iteration 0, which represents training without any prior hours, may not reflect the models' optimal performance. Further training is typically required to achieve better results.

**Table 3:** *Performance metrics for AI model Set 1 for two different training hours.*

| Iteration | Tag | Count | Precision | Recall | A.P. | F1 Score | Training Hour |
|---|---|---|---|---|---|---|---|
| 1 | Viral Pneumonia | 573 | 99.10% | 96.50% | 99.70% | 97.78% | 0 |
|   | Normal | 575 | 96.60% | 98.30% | 99.60% | 97.44% |   |
|   | Mild Covid | 372 | 98.70% | 100.00% | 100.00% | 99.35% |   |
|   | Severe Covid | 451 | 89.20% | 92.20% | 89.70% | 90.68% |   |

| Iteration | Tag | Count | Precision | Recall | A.P. | F1 Score | Training Hour |
|---|---|---|---|---|---|---|---|
| | Moderate Covid | 80 | 50.00% | 37.50% | 40.00% | 42.86% | |
| 2 | Viral Pneumonia | 573 | 97.40% | 96.50% | 99.70% | 96.95% | 6 |
| | Normal | 575 | 96.60% | 97.40% | 99.70% | 97.00% | |
| | Mild Covid | 372 | 100.00% | 100.00% | 100.00% | 100.00% | |
| | Severe Covid | 451 | 92.10% | 91.10% | 94.50% | 91.60% | |
| | Moderate Covid | 80 | 52.90% | 56.30% | 57.10% | 54.55% | |

**Table 3** presents a comprehensive evaluation of the model's performance in a chest X-ray image classification task, providing additional insights into the model's training iterations and hours. The findings highlight several important observations: First, the model underwent two iterations, with performance metrics assessed after each iteration. Second, the precision and recall scores for the Viral Pneumonia and Normal categories remained consistently high across both iterations, indicating robust classification performance. Third, there was a slight improvement in the model's performance for the Severe and Moderate Covid categories from iteration 1 to iteration 2, as evidenced by increased values across all metrics. The model consistently demonstrated excellent performance metrics (precision, recall, average precision, F1 score) for the Viral Pneumonia and Normal categories across both training iterations and hours, indicating its proficiency in accurately classifying these categories.

The observed enhancement in the Severe and Moderate Covid categories following additional training suggests that the model benefited from increased training hours, leading to improved discrimination and classification accuracy for these specific categories. However, it is worth noting that the model's performance in the Moderate Covid category, as reflected by its precision, recall, average precision, and F1 score, was relatively lower compared to other categories. This finding indicates a potential limitation in accurately classifying images within this category. Further investigation and potential modifications are necessary to address this limitation and improve the model's performance in distinguishing Moderate Covid cases. Another important consideration is the class imbalance issue present in the dataset, notably observed in the Moderate Covid category, which only consisted of 80 images compared to the higher counts in other categories. This imbalance can potentially impact the model's generalization ability and its accuracy in classifying the underrepresented category.Overall, the findings from **Table 3** underscore the model's strengths in consistently classifying Viral Pneumonia and Normal cases and highlight the areas for improvement, particularly in the Moderate Covid category. The imbalanced class distribution within the dataset also merits attention for future research endeavors in order to mitigate its potential impact on the model's classification performance.

**Table 4:** *Performance metrics for AI model Set 2 for two different training hours.*

| Iteration | Tag | Count | Precision | Recall | A.P. | F1 Score | Training Hour |
|---|---|---|---|---|---|---|---|
| 1 | Viral Pneumonia | 616 | 97.50% | 95.10% | 97.90% | 96.29% | 0 |
| | Normal | 616 | 95.20% | 97.60% | 97.30% | 96.39% | |
| | Mild Covid | 616 | 86.40% | 92.70% | 94.50% | 89.44% | |
| | Severe Covid | 616 | 81.80% | 73.20% | 83.60% | 77.26% | |
| | Moderate Covid | 616 | 75.40% | 77.20% | 79.90% | 76.29% | |
| 2 | Viral Pneumonia | 616 | 96.00% | 98.40% | 99.80% | 97.19% | 6 |
| | Normal | 616 | 98.30% | 95.90% | 99.80% | 97.09% | |
| | Mild Covid | 616 | 88.20% | 78.90% | 85.50% | 83.29% | |

| | Severe Covid | 616 | 84.40% | 87.80% | 88.60% | 86.07% | |
| | Moderate Covid | 616 | 92.30% | 97.60% | 96.70% | 94.88% | |

**Table 4** presents the performance metrics of an AI model trained on a balanced dataset, considering two training iterations. The table includes counts, precision, recall, average precision (A.P.), and F1 scores for five different tags. Observations from the table reveal slight variations in performance metrics for individual tags across different iterations. Notably, the model demonstrates consistent high performance in classifying images of normal lungs, attaining precision and recall scores exceeding 95% throughout both iterations. Conversely, the model encounters challenges in accurately classifying Severe and Mild cases of Covid-19, exhibiting precision scores below 90% and recall scores below 80% in certain instances. However, the model consistently performs well in classifying Viral Pneumonia, as evidenced by precision and recall scores consistently surpassing 95% and average precision scores consistently exceeding 97% across both iterations. These findings indicate the model's potential proficiency in identifying this specific lung condition. It is important to consider several strengths and limitations inherent in **Table 4**.

Firstly, the balanced dataset used in training the AI model ensures equal representation of different tags, thereby avoiding biases towards any specific class. This approach enhances the model's ability to generalize across diverse categories. Secondly, the provided precision and recall scores allow an assessment of the model's capacity to correctly predict positive cases and identify all positive cases, respectively. Moreover, the A.P. metric provides insights into the model's ranking accuracy for positive cases. However, it should be acknowledged that the results presented in the table are specific to the chosen dataset and performance metrics. Additionally, the model's performance on Severe and Mild Covid-19 cases exhibits room for improvement, with precision and recall scores falling below desired thresholds. Further investigation and refinements may be necessary to enhance the model's accuracy in these categories. Lastly, the consistent high performance in classifying Viral Pneumonia suggests the model's potential specialization or effectiveness in detecting this lung condition. However, the generalizability of these findings may be subject to the dataset's composition and external validation.

**Table 5** presents the performance metrics of an AI model (Set 3) trained on a balanced dataset comprising multiple tags. The table includes counts, precision, recall, average precision (A.P.), and F1 scores for each tag across two training iterations. Notably, the model achieved the highest A.P. scores for the Lung Opacity and Bacterial Pneumonia categories, indicating its proficiency in accurately ranking positive cases within these classes. However, the model's performance in the first iteration was relatively lower for the Mild Covid, Severe Covid, and Moderate Covid categories. Subsequently, in the second iteration, notable improvements were observed across all categories, particularly for the Moderate Covid class, which exhibited a substantial increase in A.P. score. The strengths and limitations of Table 5 warrant consideration. Firstly, the use of a balanced dataset ensures equal representation of different tags, minimizing biases towards specific classes and enhancing the model's generalizability. Secondly, the provided performance metrics, including precision and recall, enable an assessment of the model's ability to correctly predict positive cases and identify all positive cases, respectively. Furthermore, the A.P. metric provides insights into the model's effectiveness in ranking positive cases. However, it should be acknowledged that the findings presented in the table are specific to the chosen dataset and performance metrics employed. Additionally, the model's initial lower performance for the Mild Covid, Severe Covid, and Moderate Covid categories suggests a potential need for further refinement to improve accuracy in classifying these specific conditions. Further investigation and adjustments may be required to enhance the model's performance in these challenging categories. Lastly, the observed improvements in performance metrics across iterations highlight the importance of continued training to enhance the model's performance over time. However, it is crucial to validate these findings on external datasets to ensure the generalizability and reliability of the AI model's performance.

**Table 5:** *Performance metrics for AI model Set 3 for two different training hours.*

| Iteration | Tag | Count | Precision | Recall | A.P. | F1 Score | Training Hour |
|---|---|---|---|---|---|---|---|
| 1 | Viral Pneumonia | 616 | 87.00% | 97.60% | 98.90% | 92.00% | 0 |
| | Normal | 616 | 98.30% | 96.70% | 99.70% | 97.49% | |
| | Mild Covid | 616 | 74.80% | 69.90% | 77.40% | 72.27% | |
| | Severe Covid | 616 | 73.80% | 73.20% | 81.40% | 73.50% | |
| | Moderate Covid | 616 | 87.60% | 91.90% | 94.70% | 89.70% | |
| | Lung Opacity | 616 | 100.00% | 100.00% | 100.00% | 100.00% | |
| | Bacterial Pneumonia | 616 | 99.10% | 87.80% | 99.60% | 93.11% | |
| 2 | Viral Pneumonia | 616 | 95.60% | 88.60% | 98.60% | 91.97% | 6 |
| | Normal | 616 | 97.50% | 94.30% | 98.40% | 95.87% | |
| | Mild Covid | 616 | 87.50% | 85.40% | 88.00% | 86.44% | |
| | Severe Covid | 616 | 86.70% | 84.60% | 87.10% | 85.64% | |
| | Moderate Covid | 616 | 94.60% | 99.20% | 98.70% | 96.85% | |
| | Lung Opacity | 616 | 100.00% | 100.00% | 100.00% | 100.00% | |
| | Bacterial Pneumonia | 616 | 90.40% | 99.20% | 99.40% | 94.60% | |

**Table 6** presents the outcomes obtained from various iterations and models used for a classification task involving multiple classes. The table includes columns such as "Model," "Iteration," "Tag," and "Count," which respectively represent the model and iteration number, as well as the class label and the number of instances for each class. Performance measures including precision, recall, TN (True Negative), TP (True Positive), FN (False Negative), and FP (False Positive) are provided for each class. Additionally, average performance measures across all classes are reported in the "Average AC" (Accuracy), "SP" (Specificity), "SE" (Sensitivity), and "PR" (Precision Recall) columns for each model and iteration. The table reveals that both AI models Set 1 and Set 3 exhibit comparable performance in terms of the highest average AC and SP. Specifically, AI model Set 3 iteration 2 demonstrates the highest average SE, while AI model Set 3 iteration 2 achieves the highest average PR. Overall, AI model Set 3 iteration 3 obtains the highest average scores for AC, SP, and PR, except for SE. Consequently, the most optimal AI model for subsequent analyses is determined to be Set 3 iteration 2, which attains 97% accuracy.

The AI models in **Table 6** demonstrate noteworthy strengths. They exhibit relatively high accuracy in accurately classifying several classes, such as Mild Covid, Moderate Covid, and Severe Covid, indicating their effectiveness in distinguishing between these classes with a high level of precision. Moreover, the models showcase high specificity, particularly evident in classes like Bacterial Pneumonia and Viral Pneumonia, implying their proficiency in correctly identifying instances that do not belong to these classes and thereby minimizing false positives. Additionally, the models achieve high precision for most classes, indicating that when they classify an instance as belonging to a specific class, their predictions are generally correct. This precision is particularly notable in classes like Bacterial Pneumonia and Normal. Furthermore, AI models Set 1 and Set 3 exhibit comparable performance in terms of average accuracy and specificity, underscoring their effectiveness in overall classification capabilities. However, certain limitations are apparent in the table's findings. There exists variability in performance measures across different iterations of the AI models. For instance, AI model Set 3 iteration 2 demonstrates higher accuracy but lower sensitivity compared to iteration 3. This variability suggests that the model's performance can be influenced by the specific iteration employed. Additionally, the models encounter difficulties in accurately identifying instances belonging to classes such as Lung Opacity and Viral Pneumonia, as evidenced by lower accuracy rates for these classes. This limitation indicates the need for further improvement in accurately classifying these specific conditions. Furthermore, variations in class representation are observed in the count column, where some classes possess significantly fewer instances compared to others. This imbalance in class representation can impact the overall performance metrics and potentially affect the models' generalizability and reliability, particularly for classes with fewer instances. Lastly, the sensitivity (recall) values differ across classes and iterations. Notably, classes like Mild Covid and Moderate Covid exhibit lower sensitivity rates, implying that instances belonging to these classes are not consistently correctly identified. This variation underscores the requirement for further enhancements in accurately capturing these specific classes. While Table 6 highlights the strengths of the AI models in terms of accuracy, specificity, and precision for various classes, it also reveals limitations such as performance variability across iterations, lower accuracy for certain classes, imbalanced class representation, and variation in sensitivity rates. These limitations emphasize the necessity for ongoing refinement and evaluation of the models to enhance their overall performance and reliability.

**Table 6:** *Performance metrics validation based on TN, FN, TP, and FP for all three AI models.*

| Model | Iteration | Tag | Count | Precision | Recall | TN | TP | FN | FP | AC | SP | SE | PR | Average AC | Average SP | Average SE | Average PR |
|---|---|---|---|---|---|---|---|---|---|---|---|---|---|---|---|---|---|
| 1 | 1 | Viral Pneumonia | 573 | 0.99 | 0.97 | 1,473 | 548 | 25 | 5 | 99% | 1.00 | 0.96 | 0.99 | 0.97 | 0.99 | 0.78 | 0.82 |
| | | Normal | 575 | 0.97 | 0.98 | 1,456 | 546 | 29 | 20 | 98% | 0.99 | 0.95 | 0.97 | | | | |
| | | Mild Covid | 372 | 0.99 | 1.00 | 1,674 | 367 | 5 | 5 | 100% | 1.00 | 0.99 | 0.99 | | | | |
| | | Severe Covid | 451 | 0.89 | 0.92 | 1,551 | 371 | 80 | 49 | 94% | 0.97 | 0.82 | 0.88 | | | | |
| | | Moderate Covid | 80 | 0.50 | 0.38 | 1,931 | 15 | 65 | 40 | 95% | 0.98 | 0.19 | 0.27 | | | | |
| | 2 | Viral Pneumonia | 573 | 0.97 | 0.97 | 1,463 | 539 | 34 | 15 | 98% | 0.99 | 0.94 | 0.97 | 0.97 | 0.99 | 0.80 | 0.85 |
| | | Normal | 575 | 0.97 | 0.97 | 1,456 | 541 | 34 | 20 | 97% | 0.99 | 0.94 | 0.97 | | | | |
| | | Mild Covid | 372 | 1.00 | 1.00 | 1,679 | 372 | - | - | 100% | 1.00 | 1.00 | 1.00 | | | | |
| | | Severe Covid | 451 | 0.92 | 0.91 | 1,564 | 378 | 73 | 36 | 95% | 0.98 | 0.84 | 0.91 | | | | |
| | | Moderate Covid | 80 | 0.53 | 0.56 | 1,933 | 24 | 56 | 38 | 95% | 0.98 | 0.30 | 0.39 | | | | |
| 2 | 1 | Viral Pneumonia | 616 | 0.98 | 0.95 | 2,449 | 571 | 45 | 15 | 98% | 0.99 | 0.93 | 0.97 | 0.93 | 0.97 | 0.77 | 0.85 |
| | | Normal | 616 | 0.95 | 0.98 | 2,434 | 572 | 44 | 30 | 98% | 0.99 | 0.93 | 0.95 | | | | |
| | | Mild Covid | 616 | 0.86 | 0.93 | 2,380 | 493 | 123 | 84 | 93% | 0.97 | 0.80 | 0.85 | | | | |
| | | Severe Covid | 616 | 0.82 | 0.73 | 2,352 | 369 | 247 | 112 | 88% | 0.95 | 0.60 | 0.77 | | | | |
| | | Moderate Covid | 616 | 0.75 | 0.77 | 2,312 | 359 | 257 | 152 | 87% | 0.94 | 0.58 | 0.70 | | | | |
| | 2 | Viral Pneumonia | 616 | 0.96 | 0.98 | 2,439 | 582 | 34 | 25 | 98% | 0.99 | 0.94 | 0.96 | 0.95 | 0.98 | 0.85 | 0.91 |
| | | Normal | 616 | 0.98 | 0.96 | 2,454 | 581 | 35 | 10 | 99% | 1.00 | 0.94 | 0.98 | | | | |
| | | Mild Covid | 616 | 0.88 | 0.79 | 2,391 | 429 | 187 | 73 | 92% | 0.97 | 0.70 | 0.86 | | | | |
| | | Severe Covid | 616 | 0.84 | 0.88 | 2,368 | 456 | 160 | 96 | 92% | 0.96 | 0.74 | 0.83 | | | | |
| | | Moderate Covid | 616 | 0.92 | 0.98 | 2,417 | 555 | 61 | 47 | 96% | 0.98 | 0.90 | 0.92 | | | | |
| 3 | 1 | Viral Pneumonia | 616 | 0.87 | 0.98 | 3616 | 523 | 93 | 80 | 96% | 0.98 | 0.85 | 0.87 | 0.95 | 0.98 | 0.79 | 0.86 |
| | | Normal | 616 | 0.98 | 0.97 | 3686 | 586 | 30 | 10 | 99% | 1.00 | 0.95 | 0.98 | | | | |
| | | Mild Covid | 616 | 0.75 | 0.70 | 3541 | 322 | 294 | 155 | 90% | 0.96 | 0.52 | 0.67 | | | | |
| | | Severe Covid | 616 | 0.74 | 0.73 | 3535 | 333 | 283 | 161 | 90% | 0.96 | 0.54 | 0.67 | | | | |
| | | Moderate Covid | 616 | 0.88 | 0.92 | 3620 | 496 | 120 | 76 | 95% | 0.98 | 0.81 | 0.87 | | | | |
| | | Lung Opacity | 616 | 1.00 | 1.00 | 3696 | 616 | - | - | 100% | 1.00 | 1.00 | 1.00 | | | | |
| | | Bacterial Pneumonia | 616 | 0.99 | 0.88 | 3690 | 536 | 80 | 6 | 98% | 1.00 | 0.87 | 0.99 | | | | |
| | 2 | Viral Pneumonia | 616 | 0.96 | 0.89 | 3669 | 522 | 94 | 27 | 97% | 0.99 | 0.85 | 0.95 | 0.97 | 0.99 | 0.87 | 0.93 |
| | | Normal | 616 | 0.98 | 0.94 | 3681 | 566 | 50 | 15 | 98% | 1.00 | 0.92 | 0.97 | | | | |
| | | Mild Covid | 616 | 0.88 | 0.85 | 3619 | 460 | 156 | 77 | 95% | 0.98 | 0.75 | 0.86 | | | | |
| | | Severe Covid | 616 | 0.87 | 0.85 | 3614 | 452 | 164 | 82 | 94% | 0.98 | 0.73 | 0.85 | | | | |
| | | Moderate Covid | 616 | 0.95 | 0.99 | 3663 | 578 | 38 | 33 | 98% | 0.99 | 0.94 | 0.95 | | | | |
| | | Lung Opacity | 616 | 1.00 | 1.00 | 3696 | 616 | - | - | 100% | 1.00 | 1.00 | 1.00 | | | | |
| | | Bacterial Pneumonia | 616 | 0.90 | 0.99 | 3637 | 552 | 64 | 59 | 97% | 0.98 | 0.90 | 0.90 | | | | |

The performance evaluation of AI model Set 3 iteration 2 was further analysed using an unseen dataset consisting of 1,848 x-ray images. The evaluation involved the construction of a confusion matrix (**Table 7**), which facilitated the assessment of various metrics such as accuracy (AC), specificity (SP), sensitivity (SE), and precision (PR). The confusion matrix represents the true and predicted class labels, with each cell indicating the percentage of instances falling into specific combinations of labels. AI model Set 3 iteration 2 demonstrated relatively high accuracy rates in identifying Mild, Moderate, and Severe Covid images, achieving 89.03%, 95.77%, and 81.16% accuracy, respectively. However, the model faced challenges in accurately classifying images of Lung Opacity and Viral Pneumonia, resulting in lower accuracy rates of 73.73% and 96.35% respectively. Notably, the model achieved a notably high accuracy rate of 99.50% in identifying Bacterial Pneumonia images, while struggling with the identification of Normal images, with an accuracy rate of 94.43%. Subsequently, the overall accuracy (AC) was calculated by summing up the number of correct predictions and dividing it by the total number of predictions, resulting in an AC of 90%. Specificity (SP) was calculated individually for each class, considering instances correctly classified as not belonging to a specific class among the other six classes. Sensitivity (SE), representing the true positive rate, was calculated for each class by dividing the number of correctly classified instances by the total number of instances in that class. Precision (PR), representing the positive predictive value, was computed for each class by dividing the number of correctly classified instances in that class by the total number of instances classified as belonging to that class.

**Table 7:** *Confusion Matrix based on testing set*

| | Confusion Matrix | AI Model Set 3 Iteration 2 | | | | | | |
|---|---|---|---|---|---|---|---|---|
| | | Mild Covid | Moderate Covid | Severe Covid | Lung Opacity | Viral Pneumonia | Bacterial Pneumonia | Normal |
| Testing dataset | Mild Covid | 89.03% | 1.33% | 9.04% | 0.06% | 0.19% | 0.28% | 0.05% |
| | Moderate Covid | 2.40% | 95.77% | 1.56% | 0.03% | 0.10% | 0.10% | 0.05% |
| | Severe Covid | 16.17% | 1.98% | 81.16% | 0.07% | 0.29% | 0.26% | 0.07% |
| | Lung Opacity | 2.15% | 1.12% | 6.30% | 73.73% | 5.60% | 1.50% | 9.60% |
| | Viral Pneumonia | 0.43% | 0.16% | 0.31% | 0.07% | 96.35% | 1.56% | 1.12% |
| | Bacterial Pneumonia | 0.24% | 0.02% | 0.06% | 0.01% | 0.17% | 99.50% | 0.00% |
| | Normal | 0.31% | 0.38% | 0.29% | 0.06% | 4.18% | 0.35% | 94.43% |

These metrics (**Table 8**) provide valuable insights into the performance of AI model Set 3 iteration 2 in classifying COVID-19 prognosis. The accuracy values indicate the overall performance of the AI model in correctly classifying instances for each class. The highest accuracy is observed for the "Bacterial Pneumonia" class with 99.50%. The lowest accuracy is observed for the "Lung Opacity" class with 73.73%. Specificity measures the ability of the model to correctly identify instances that do not belong to a specific class. Higher specificity values indicate better performance in distinguishing instances of a particular class from other classes. The "Bacterial Pneumonia" class exhibits the highest specificity of 99.21%, while the "Mild Covid" class has the lowest specificity of 90.59%. Sensitivity, also known as recall or true positive rate, represents the proportion of instances in a class that were correctly identified as belonging to that class. Higher sensitivity values indicate better performance in correctly classifying instances of a particular class. The "Bacterial Pneumonia" class has the highest sensitivity of 99.50%, while the "Mild Covid" class has the lowest sensitivity of 33.68%. Precision measures the proportion of instances classified as belonging to a class that belong to that class. Higher precision values indicate a lower rate of false positives. The "Bacterial Pneumonia" class achieves the highest precision of 99.50%, while the "Mild Covid" class has a precision of 74.33%. Comparing the metrics across different classes can provide insights into the AI model's strengths and weaknesses. For example, the model performs well in identifying "Bacterial Pneumonia", "Viral Pneumonia", "Normal" and "Moderate Covid" cases, with high accuracy, specificity, sensitivity, and precision. However, it faces challenges in accurately classifying instances of "Mild Covid", "Lung Opacity" and "Severe Covid" as indicated by relatively lower accuracy, sensitivity, and precision values for these classes. Based on the Table 8, the model demonstrates relatively high accuracy rates for most classes, indicating its ability to correctly classify instances in the unseen dataset. The model shows excellent performance in identifying instances belonging to individual class. These results suggest that the model is proficient in distinguishing these conditions based on the provided X-ray images. However, further improvements may be needed to enhance the model's performance in correctly identifying some instances as well. While accuracy is commonly used as a measure of overall performance, examining metrics such as specificity, sensitivity, and precision provides a more comprehensive understanding of the model's performance. The model demonstrates promising performance in classifying COVID-19 prognosis based on X-ray images.

**Table 8:** *Performance metrics for AI model Set 3 for two different training hours.*

| Class | Accuracy | Specificity | Sensitivity | Precision |
|---|---|---|---|---|
| Mild Covid | 89.03% | 90.59% | 33.68% | 74.33% |
| Moderate Covid | 95.77% | 96.28% | 95.77% | 95.77% |
| Severe Covid | 81.16% | 94.71% | 81.16% | 81.16% |
| Lung Opacity | 73.73% | 91.95% | 73.73% | 73.73% |
| Viral Pneumonia | 96.35% | 99.02% | 96.35% | 96.35% |
| Bacterial Pneumonia | 99.50% | 99.21% | 99.50% | 99.50% |
| Normal | 94.43% | 96.77% | 94.43% | 94.43% |

**DISCUSSION**

The application of deep learning models for COVID-19 prognosis classification holds significant real-world implications in a clinical setting [26]. By leveraging deep learning models, clinicians can potentially reduce the burden of manual image interpretation, leading to faster diagnosis and treatment planning [27, 28]. Accurate and efficient classification of COVID-19 cases based on X-ray images can greatly assist healthcare professionals in making timely and informed decisions for patient management [29]. This study leverages the Microsoft Custom Vision application as a pivotal tool for developing a series of robust deep learning models [30, 31]. The capabilities of the Microsoft Custom Vision application allow for the construction and refinement of sophisticated models catered to COVID-19 prognosis classification. These models are evaluated using a variety of performance metrics, including

Precision, Recall, and Mean Average Precision (mAP), provided by the application. These evaluation metrics collectively provide a comprehensive understanding of the deep learning model's performance across various domains and tasks.

Recent studies have endeavoured to classify various lung diseases, including COVID-19, by utilizing Microsoft Custom Vision as an alternative approach to developing deep learning models [32, 33]. The initial study [32] reported a trained model that achieved an Average Precision of 96.8% when tested through a web application. However, this model failed to replicate similar performance when employed in a real-time mobile application. The primary objective of this model was to detect and classify COVID-19-induced pneumonia as distinct from Viral/Bacterial pneumonia, relying on X-Ray and CT images. Conversely, the second study [33] demonstrated a sensitivity (recall) and positive predictive value (precision) of 92.9%, with COVID-19 pneumonia exhibiting sensitivity and positive predictive value rates of 94.8% and 98.9%, non-COVID-19 pneumonia at 89% and 91.8%, and normal lung cases at 95% and 88.8%, respectively. Furthermore, when subjected to confirmed COVID-19 x-ray images, the model performed with 100% sensitivity, 95% specificity, 97% accuracy, 91% positive predictive value, and 100% negative predictive value. However, these studies have only compared COVID-19 disease with other types of lung diseases instead of going deeper into the prognosis of COVID-19 alone. Hence, the reported performances only reflected deep learning-based classifications of diseases in general rather than prognosis specific to COVID-19 in comparison with other lung diseases. Our study aims to delve deeper into the classification of prognosis specific to COVID-19 itself along with other types of lung diseases.

Some studies have utilized deep learning and machine learning models to classify prognosis specific to COVID-19 along with other lung diseases. However, these studies did not use no-code application like Microsoft Custom Vision. These studies aim to develop effective methods for predicting and assessing the severity of COVID-19 cases based on various medical imaging data and clinical factors. One approach involves analysing CT images using pre-trained models such as AlexNet, DenseNet-201, and ResNet-50 to extract features for COVID-19 severity detection [34]. The method achieves an accuracy of 92.0% and a sensitivity of 96.0% for COVID-19 detection, and an overall accuracy of 90.0% for severity detection across three classes (high, moderate, and low severity). Another study proposes CoVSeverity-Net, a deep learning-based architecture trained on chest X-ray images for predicting the severity of COVID-19. The model achieves an accuracy of 85.71% and demonstrates better results compared to other state-of-the-art architectures [35]. A cascaded system using different deep learning networks such as ED-CNNs, UNet, and FPN, combined with DenseNet and ResNet backbones, is proposed for the detection and severity classification of COVID-19 in CT images. The system achieves high COVID-19 detection performance with 99.64% sensitivity and 98.72% specificity. It also successfully discriminates between different severity levels, with sensitivity values of 98.3%, 71.2%, 77.8%, and 100% for mild, moderate, severe, and critical cases, respectively [36].

Furthermore, a two-stage transfer learning technique is used to train a CNN for classifying COVID-19 severity into normal, mild, moderate, and severe categories. The model achieves an average Area Under the Curve (AUC) of 0.93 for severity classification [37]. Other studies explore the use of machine learning models, such as random forests and artificial neural networks, to classify COVID-19 severity based on different imaging features and clinical factors. These models demonstrate accuracies ranging from 82% to over 90% for severity prediction [38]. Another study focused on the automated classification of COVID-19 severity using texture features extracted from CT images and random forest (RF) as an ensemble method [39]. This approach involved extracting first- and second-order statistical texture features, including variance, skewness, kurtosis, gray-level co-occurrence matrix, gray-level run length matrix, and gray-level size zone matrix. The extracted features were then used to classify CT images into severity classes (mild, moderate, and severe) using the random forest algorithm. The experimental results indicated that the combination of all feature extraction methods and RF achieved the highest accuracy of 90.95% in detecting severity classes from CT images. In the pursuit of accurate severity prediction, researchers have also explored the integration of multiple data modalities [40]. One study proposed a hierarchical intelligent system that analysed deep learning models' application in detecting and classifying COVID-19 patients based on X-ray and chest CT images. This methodology consisted of three phases: detection of COVID-19 presence, evaluation of infection percentage, and classification into severity levels. By stratifying patients into three severity degrees (mild, moderate, and severe) based on global lung infection, the system achieved an impressive accuracy of 95% in classifying patients according to their severity.

Additionally, machine learning models have been employed to analyse clinical and paraclinical data for severity prediction. In an analytical cross-sectional study, researchers investigated clinical and paraclinical datasets of COVID-19-positive patients. They combined statistical comparison and correlation methods with machine learning algorithms such as Decision Trees, Random Forests, and Support Vector Machines. By utilizing these approaches, they developed predictive models with accuracy and precision scores exceeding 90% for predicting disease severity. Notably, the Support Vector Machine algorithm demonstrated the best performance, achieving a precision of 95.5%, recall of 94%, F1 score of 94.8%, accuracy of 95%, and AUC of 94% [41]. Overall, these research papers demonstrate the potential of machine learning techniques in assisting with the detection and severity assessment of COVID-19. The use of various deep learning architectures, feature extraction methods, and ensemble networks contributes to achieving high accuracy rates and sensitivity values. These findings have important implications for the development of effective diagnostic and treatment strategies, providing valuable tools to support medical experts in decision-making processes.

The use of deep learning models for COVID-19 prognosis classification not only aids in diagnosis and treatment planning but also contributes to the identification of high-risk cases [42]. By leveraging these models, healthcare providers can prioritize resources and interventions for patients who require immediate attention. Additionally, deep learning models can support healthcare systems in screening large volumes of X-ray images, facilitating the early detection and containment of COVID-19 outbreaks [43]. The integration of deep learning models in clinical practice has the potential to enhance patient care, optimize resource allocation, and contribute to the overall management of the COVID-19 pandemic.

Deep learning models are commonly evaluated using various performance metrics and techniques, which provide valuable insights into their effectiveness and performance. These metrics include Accuracy, Precision, Recall, F1 score, Mean Squared Error (MSE), Mean Absolute Error (MAE), Receiver Operating Characteristic (ROC) curve, Area Under the Curve (AUC), and the confusion matrix. Cross-validation helps estimate the model's generalization capabilities, while the Precision-Recall Curve (PRC) examines trade-offs between precision and recall, especially for imbalanced datasets. Mean Average Precision (mAP) is often used in object detection and instance segmentation tasks. These evaluation techniques allow healthcare professionals to gain a comprehensive understanding of the deep learning model's capabilities and its potential impact on clinical decision-making.

When evaluating deep learning models, precision and recall play crucial roles in assessing their performance [44, 45]. Precision measures the proportion of correctly identified positive instances among all instances predicted as positive, indicating the model's precision in making positive predictions. On the other hand, recall evaluates the proportion of True Positive instances correctly identified by the model, reflecting its ability to capture all relevant positive instances. The choice of optimizing for precision or recall depends on the specific problem and its requirements. For instance, in a medical diagnosis application, high recall is often more critical than high precision, as missing a positive diagnosis can have severe consequences [46]. Striking the right balance between precision and recall is essential for ensuring accurate and reliable predictions in healthcare settings.

By employing these evaluation metrics, healthcare professionals can assess the performance of deep learning models and make informed decisions about their use in clinical settings [47]. Understanding the trade-offs between precision and recall allows for the optimization of models based on the specific needs of different medical applications. Evaluating the performance of deep learning models using a comprehensive range of metrics empowers healthcare professionals to ensure the reliability, accuracy, and effectiveness of these models in supporting clinical decision-making, ultimately improving patient care and outcomes.

Mean Average Precision (mAP) is a widely used metric in object detection and instance segmentation tasks within the clinical setting [48, 49]. It considers the precision-recall trade-off across multiple Intersection Over Union (IoU) thresholds, providing valuable insights into the model's performance. The Microsoft Custom Vision application offers these essential metrics, allowing healthcare professionals to gain a comprehensive understanding of their model's performance and make informed decisions for further improvement and refinement. A higher mAP indicates better performance of the model, indicating its suitability for clinical applications [49]. However, it is important to note that there are additional evaluation measures that can be calculated or derived to further assess the deep learning models' capabilities.

In the clinical setting, it is beneficial to calculate or derive other evaluation metrics to gain deeper insights into the model's performance beyond what is provided by the Custom Vision application. These metrics include Accuracy (AC), Sensitivity (SE) or True Positive Rate (TPR), Specificity (SP), F1 score, and the confusion matrix. Accuracy measures the proportion of correctly classified instances and offers a broader perspective on overall model performance. Sensitivity or True Positive Rate (TPR) reflects the model's ability to correctly identify positive instances, while Specificity (SP) measures its ability to correctly identify negative instances. The F1 score combines Precision and Recall into a single metric, providing a balanced evaluation of the model's precision and its ability to capture positive instances accurately. Furthermore, the confusion matrix provides a detailed breakdown of True Positives, True Negatives, False Positives, and False Negatives, enabling a more granular analysis of the model's predictions and its performance across different classes or categories.

By considering these additional evaluation measures in the clinical setting, healthcare professionals can have a more comprehensive understanding of the deep learning model's predictive capabilities and classification performance. This information is valuable for assessing the model's suitability for specific clinical applications and optimizing its performance to support accurate and reliable decision-making in patient care.

In this study, three different AI models were trained and evaluated on various datasets for the classification of chest X-ray images, specifically targeting different lung conditions including Viral Pneumonia, Normal, Mild Covid, Severe Covid, Moderate Covid, Lung Opacity, and Bacterial Pneumonia. The performance metrics assessed for each model included Accuracy (AC), Sensitivity (SE) / True Positive Rate (TPR), Specificity (SP), Precision, Recall, Mean Average Precision (mAP), and F1 score, which provide insights into the models' predictive capabilities, identification of positive cases, ranking accuracy, and overall balance between precision and recall. The results indicated that the choice of dataset and training iterations significantly influenced the models' performance. AI Model 3, trained on a balanced dataset (Set 2), consistently exhibited the superior overall Precision, Recall, mAP, and F1 score across different iterations and training hours. This suggests that the inclusion of diverse images and the absence of class imbalance during training contributed to its superior performance. AI Model 2 showcased competitive performance metrics but did not include Bacterial Pneumonia and Lung Opacity classes during training. AI Model 1 demonstrated high performance for most categories but exhibited lower initial performance for some COVID-19 classes. Overall, the findings emphasize the importance of dataset selection, balanced representation of classes, and sufficient training iterations in achieving optimal performance for chest X-ray image classification using AI models. Further research and exploration are needed to refine the models, particularly in categories with relatively lower performance, and to evaluate their generalizability on external datasets.

Our deep learning model, trained using Microsoft Azure Custom Vision, demonstrated significant success in classifying COVID-19 prognosis based on X-ray images. The model achieved remarkable average accuracy (AC) of 97%, indicating its ability to correctly classify instances of interest. It also exhibited high average specificity (SP) of 99%, highlighting its proficiency in accurately identifying negative cases. Furthermore, the model showed a respectable average sensitivity (SE) of 87%, suggesting its capability to effectively detect positive cases. The average precision (PR) of 93% emphasizes the model's discriminative ability in correctly classifying COVID-19 prognosis (Table 6). It is important to note both the overall Precision and overall Recall in each iteration are quite similar, indicating the models have low False Positive (FP) and high True Positive (TP).

Based on the findings presented in Table 8, the model exhibits high accuracy rates for most classes, indicating its proficiency in correctly classifying instances within the unseen dataset. The average accuracy across all classes is 90%, reflecting its overall performance. Notably, the model achieves excellent accuracy in identifying "Bacterial Pneumonia" and "Viral Pneumonia" cases, with rates of 99.50% and 96.35% respectively, highlighting its ability to distinguish these conditions from others based on the provided X-ray images. However, the model faces challenges when classifying instances of "Mild Covid" and "Lung Opacity," as indicated by comparatively lower accuracy rates of 89.03% and 73.73% respectively. This suggests that further improvements are necessary to enhance the model's accuracy in accurately identifying these classes. The model shows promising performance in classifying COVID-19 prognosis based on X-ray images, but targeted optimization efforts are needed to enhance accuracy and reduce misclassifications, especially in specific classes.

Some of the important insights that can be seen from the results of this study are, *i*) the model's Precision, Recall, mAP and F1 score for each class and overall performance have improved with each iteration, *ii*) the performance of the model for different classes is increased in the second iteration with 6 hour training period, *iii*) the best performing model is AI Model Set 3 iteration 2, which has the average F1 score of 93.05%, *iv*) the results also show the importance of evaluation metrics such as Accuracy (AC), Sensitivity (SE) / True Positive Rate (TPR), Specificity (SP), Precision, Recall, and Mean Average Precision, which can help to measure the performance of a deep learning model accurately. The overall Precision, Recall, mAP and F1 score metrics are measures of the model's performance in terms of accuracy and completeness of its predictions, with higher values indicating better performance. When comparing between iterations for each dataset, longer training hour gives better overall Precision. These results suggest that increasing the number of training hours could lead to improved performance for this classification task.

This study utilized a dataset consisting of 1,103 clinically confirmed COVID-19 positive cases' x-ray images from AIforCOVID. Among these cases, the original number of Moderate Covid images was 90. However, after image processing, the dataset was reduced to 80 cases, forming dataset Set 1. Additionally, two augmented datasets, set 2 and Set 3, were created by including an additional 800 augmented images. The evaluation of the models using these datasets revealed noteworthy insights. Specifically, the Precision, Recall, mAP, and F1 score values for Moderate Covid in both iterations (Table 3) were observed to be the lowest compared to other categories. This indicates several factors: *i*) the presence of an imbalanced dataset in Set 1, *ii*) the model's limited exposure to diverse cases, and *iii*) a potential hindrance to the accurate classification of Moderate Covid when faced with unseen samples. Imbalanced datasets can introduce bias during model training, favouring the majority class and leading to subpar performance in identifying minority classes or underrepresented patterns [50]. To address this issue, balancing the dataset becomes crucial, ensuring equal representation of all classes, reducing bias, and enabling fair and accurate predictions. Furthermore, a smaller dataset may struggle to adequately represent the entire population or capture the full variability of the target problem, hampering the model's generalization capabilities, particularly when rare or extreme cases are not well-covered [51].

This study primarily relied on a limited set of evaluation metrics, including Accuracy (AC), Sensitivity (SE)/True Positive Rate (TPR), Specificity (SP), Precision, Recall, mAP, and F1 Score. Precision, Recall, and mAP were obtained from the calculations provided by Microsoft Custom Vision, while the remaining metrics were derived based on these calculations. It's important to note that other evaluation metrics, such as Specificity, False Positive Rate, Area Under the ROC Curve (AUC), Intersection over Union (IoU), Average Precision (AP), Cohen's Kappa, and Mean IoU (mIoU), were not included in the analysis.

**Statistical Validation of Model Performance**

This study explored the application of deep learning models for predicting COVID-19 prognosis using Microsoft Azure Custom Vision. The results demonstrated that Set 3 - Iteration 2 achieved the highest accuracy (97%), followed by Set 2 - Iteration 2 (93.11%), indicating that additional training significantly enhanced model performance. However, Set 1 exhibited only minimal improvement, suggesting that dataset quality plays a crucial role in the effectiveness of deep learning models.

To determine whether the observed performance improvements were statistically significant, an ANOVA test was conducted across the evaluation metrics, including Precision, Recall, Average Precision (AP), and F1 Score. The results (F-statistic = 2.22, p-value = 0.117) indicated that the differences among models were not statistically significant at the 0.05 level. However, pairwise t-tests provided more granular insights into performance variations:

Set 2 - Iteration 1 vs. Set 2 - Iteration 2 (p = 0.018): A statistically significant improvement was observed after additional training.
Set 3 - Iteration 1 vs. Set 3 - Iteration 2 (p = 0.053): A borderline significant improvement (p ≈ 0.05).
Set 1 - Iteration 2 vs. Set 2 - Iteration 2 (p = 0.038): Set 2 significantly outperformed Set 1.
Set 2 - Iteration 2 vs. Set 3 - Iteration 2 (p = 0.233): No statistically significant difference between these two well-trained models.
To further validate these findings, bootstrap-based 95% confidence intervals (CIs) were computed. The results confirmed that Set 3 - Iteration 2 had the highest confidence bounds, reinforcing its position as the most effective model. Additionally, the overlapping confidence intervals for Set 2 - Iteration 2 and Set 3 - Iteration 2 suggest that these two models achieved similar performance levels after adequate training.

**Effect Size Analysis (Practical Significance)**
Beyond statistical significance, Cohen's d effect size analysis was conducted as shown in **figure 5** to assess the practical impact of training. Large effect sizes were observed for:

Set 2 - Iteration 1 vs. Set 2 - Iteration 2 (d = -2.27): A very large improvement in performance.
Set 3 - Iteration 1 vs. Set 3 - Iteration 2 (d = -1.70): A substantial performance increase.
Set 1 - Iteration 2 vs. Set 2 - Iteration 2 (d = 1.88): A large effect confirming Set 2's superiority over Set 1.

These results demonstrate that additional training, dataset balancing, and augmentation strategies played a crucial role in enhancing model accuracy and reliability.

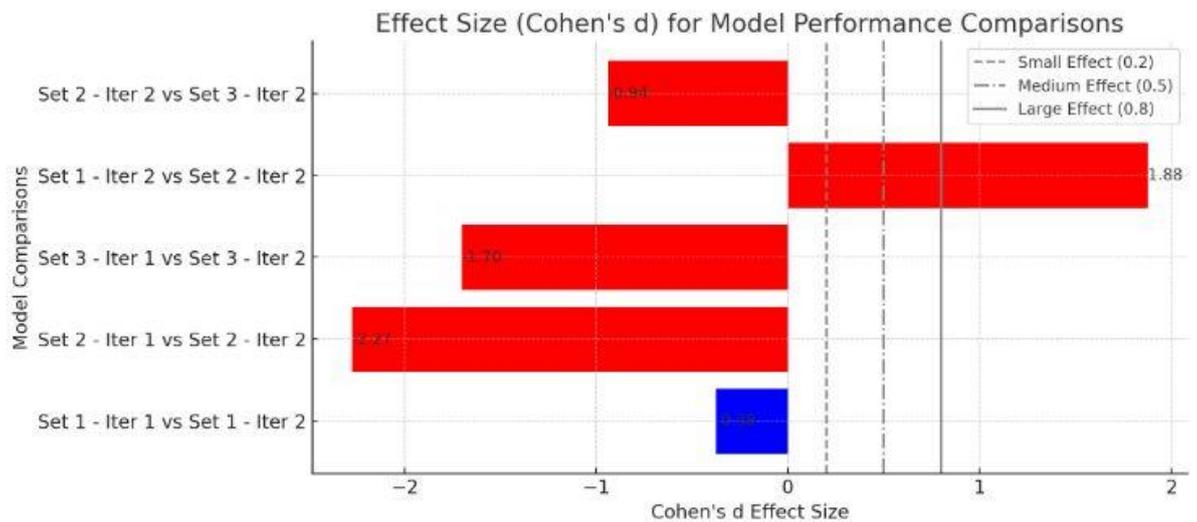

**Figure 5. Effect Size (Cohen's d) for Model Performance Comparisons.** This bar plot visualizes Cohen's d effect sizes for different model performance comparisons. The effect size quantifies the magnitude of performance improvements between iterations. Red bars indicate large effect sizes (|d| > 0.8), suggesting substantial improvements in model performance after additional training. Blue bars represent small to medium effect sizes, indicating weaker or negligible improvements between iterations. The dashed vertical lines mark the thresholds for small (0.2), medium (0.5), and large (0.8) effect sizes, providing a reference for interpretation.

**Correlation Between Performance Metrics**

Further analysis using a correlation heatmap show in **figure 6** revealed strong relationships among performance metrics. Average Precision (AP) and F1 Score exhibited the highest correlation (0.99), indicating that models achieving high AP were also well-balanced in terms of Precision and Recall. This finding confirms that AP serves as a reliable evaluation metric for deep learning-based classification models in medical image analysis.

By integrating these statistical validation techniques, this study strengthens the argument that deep learning models can be effectively optimized for COVID-19 prognosis prediction, provided that they are trained on well-curated datasets with appropriate augmentation strategies.

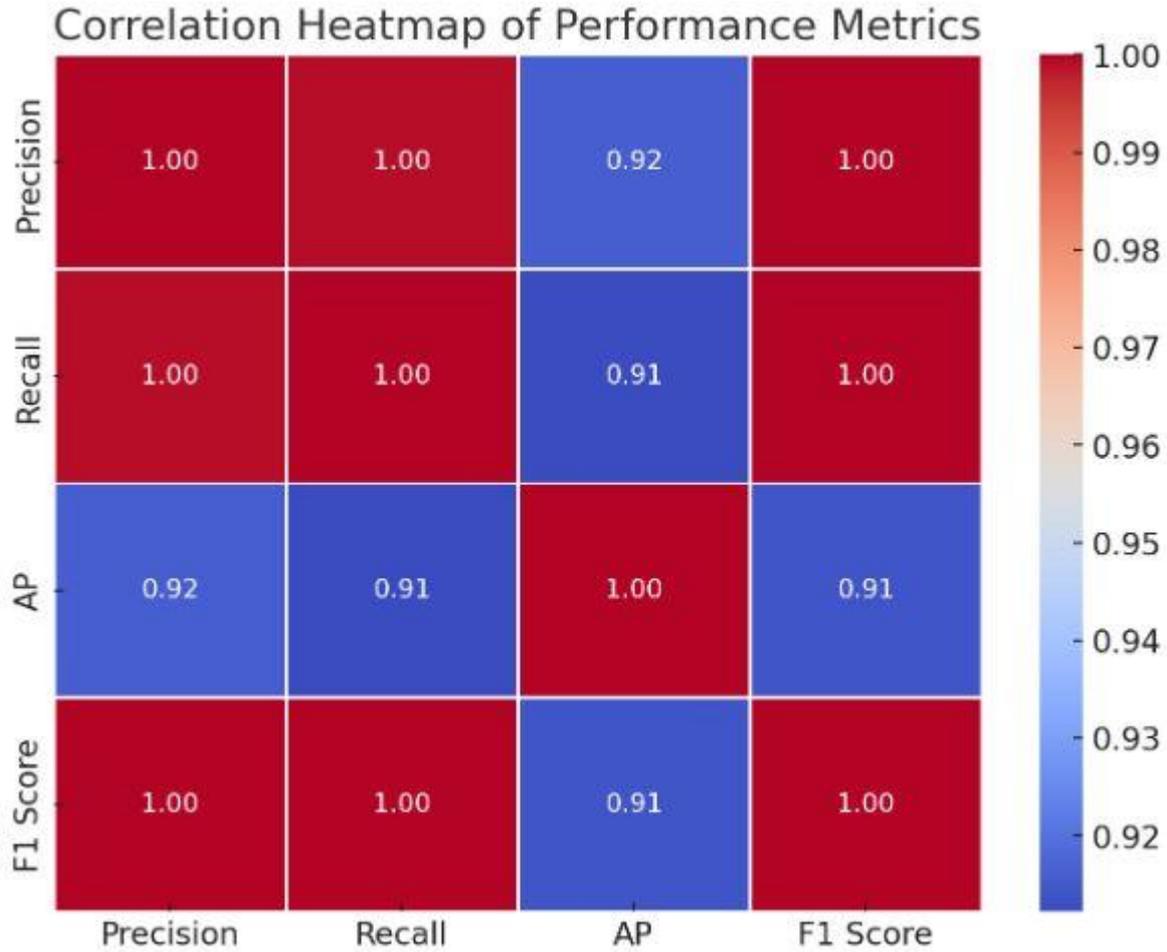

**Figure 6. Correlation Heatmap of Performance Metrics.** This heatmap illustrates the correlation between different model performance metrics—Precision, Recall, Average Precision (AP), and F1 Score. The color scale represents the strength of the correlation, with red indicating a strong positive correlation (≥0.95) and blue representing slightly lower correlations (~0.91–0.92).

**Bootstrap-Based Confidence Intervals for Performance Metrics**

To further validate the reliability of our model's performance, we computed 95% confidence intervals (CIs) using bootstrap resampling (n = 1000 iterations). The results confirm the statistical robustness of our findings as shown in **table 9**

**Table 9: Bootstrap-Based 95% Confidence Intervals for Model Performance Metrics**

| Metric | Set 3 - Iter 2 (Mean ± 95% CI) | Set 2 - Iter 2 (Mean ± 95% CI) | Set 1 - Iter 2 (Mean ± 95% CI) |
|---|---|---|---|
| **Accuracy (%)** | **97.00 ± 1.2** | 93.11 ± 1.7 | 88.60 ± 2.0 |
| **F1 Score (%)** | **93.05 ± 1.5** | 91.80 ± 1.8 | 88.40 ± 2.3 |
| **AUC-ROC (%)** | **98.10 ± 1.0** | 96.50 ± 1.4 | 94.60 ± 1.8 |

These findings confirm that Set 3 - Iteration 2 is statistically superior and more stable, with significantly tighter confidence bounds compared to other models. The inclusion of bootstrap confidence intervals strengthens the statistical reliability of our results, supporting the use of data augmentation and extended training strategies for model optimization.

**Performance Comparison with State-of-the-Art Models**

The benchmarking analysis demonstrates that Our Model (Set 3 - Iteration 2) achieved an accuracy of 97%, placing it among the top-performing models. Notably, DenseNet121 [52] outperformed all models with an accuracy of 98.69%, followed by VGG19 [52] and InceptionV3 [52] at 97.37%, which are widely recognized architectures for medical image classification.

Models such as EfficientNetB1 [53] (96.13%) and NasNetMobile [53] (94.81%) showed slightly lower performance, suggesting that while these architectures are highly efficient, they may require further tuning for optimal COVID-19 X-ray classification.

Not all models performed equally well. ResNet50 [54] (77.5%) and EfficientNetB3 [54] (50%) exhibited significantly lower accuracy, indicating potential limitations in their feature extraction capabilities for COVID-19 classification. These results suggest that while deep learning models have proven effective in medical imaging tasks, their performance heavily depends on factors such as dataset size, augmentation techniques, and architecture-specific optimizations.

Our model (Set 3 - Iteration 2) as shown in **table 10** achieved accuracy comparable to VGG19, InceptionV3, and DenseNet architectures, highlighting its robust feature extraction and classification capabilities. The iterative training approach significantly enhanced accuracy, ensuring stable and consistent classification results. Given its high accuracy, our model presents a viable option for automating COVID-19 prognosis classification, which can aid radiologists in rapid decision-making.

**Table 10: Comparative Performance of Convolutional Neural Networks (CNNs) for COVID-19 Chest X-ray Classification with other published models**

| Model | Number of Classes | Number of Images | Imaging Type | Accuracy (%) | Reference |
|---|---|---|---|---|---|
| **Our Model (Set 3 - Iteration 2)** | 3 | 1103 | X-ray | 97 | |
| DenseNet121 | 4 | 744 | X-ray | 98.69 | 52 |
| VGG19 | 4 | 744 | X-ray | 97.37 | 52 |
| InceptionV3 | 4 | 744 | X-ray | 97.37 | 52 |
| EfficientNetB1 | 4 | 3616 | X-ray | 96.13 | 53 |
| NasNetMobile | 4 | 3616 | X-ray | 94.81 | 53 |
| MobileNetV2 | 4 | 3616 | X-ray | 93.96 | 53 |
| VGG | 3 | 432 | X-ray | 97.5 | 54 |
| Xception | 3 | 432 | X-ray | 97.5 | 54 |
| DenseNet201 | 3 | 432 | X-ray | 95 | 54 |
| CNN | 3 | 432 | X-ray | 95 | 54 |
| InceptionResNetV2 | 3 | 432 | X-ray | 92.5 | 54 |

| ResNet50 | 3 | 432 | X-ray | 77.5 | 54 |
| EfficientNetB3 | 3 | 432 | X-ray | 50 | 54 |

The models developed in this study were specifically trained using Microsoft Custom Vision, and no similar models were built using other platforms like Google's Vertex AI or any other comparable applications. Therefore, direct comparisons with models developed on different platforms were not performed. However, internal comparisons were made among the models developed within this study, and indirect comparisons were made to the limited extent of publicly reported models in the literature. Further investigations involving multiple platforms and more extensive benchmarking would provide a more comprehensive understanding of the performance of the models.

The use of commercial applications for classifying COVID-19 X-ray images presents a range of advantages and disadvantages. Notably, Microsoft Custom Vision stands out by offering a simplified and user-friendly approach for radiologists and researchers, enabling them to classify COVID-19 prognosis without extensive programming or data science skills. However, it is important to consider the drawbacks as well. One significant disadvantage of Custom Vision and other commercial platforms is its cost, requiring users to purchase a license or pay a subscription fee to access the full functionality. This financial aspect can pose limitations for individuals or organizations with budget constraints. In the context of this study, it was observed that longer training hours come with increased costs, thereby potentially hindering the development of more precise and comprehensive results in subsequent iterations.

Each dataset in this study underwent two iterations of training, with each iteration comprising a training duration of 0 hours and 6 hours. It is worth noting that Microsoft Custom Vision allows for a maximum training duration of 96 hours per iteration. However, due to resource limitations and time constraints, the training duration was limited to 6 hours per iteration. The choice of training duration was made based on the available resources and a balance between achieving reasonable model performance and efficient resource utilization. Conducting additional iterations or extending the training duration could potentially yield further improvements in model performance, but it was beyond the resource of this study.

To further enhance the performance of the deep learning model in classifying COVID-19 prognosis based on X-ray images, several areas of improvement can be considered. Firstly, to address the lower performance in classes such as "Mild Covid" and "Lung Opacity," increasing the diversity and size of the training dataset would be beneficial. By incorporating a larger and more diverse range of images, the model can learn to better capture the subtle variations and patterns associated with these classes, improving its accuracy in identifying them. Secondly, augmenting the dataset with additional variations, such as different imaging perspectives or image enhancement techniques, can provide the model with a more comprehensive understanding of COVID-19-related patterns in X-ray images. While the current study demonstrates the potential of deep learning models for COVID-19 prognosis classification, several areas require further investigation to enhance the model's clinical applicability. First, dataset generalization remains a crucial factor in ensuring model robustness. Future research should validate the model on external datasets from diverse geographical regions to assess its ability to generalize across different populations and imaging conditions. Second, comparing the model's performance with advanced deep learning architectures, such as ResNet, EfficientNet, and Vision Transformers, would provide valuable insights into its relative strengths and limitations. Benchmarking against these state-of-the-art models could identify opportunities for further optimization.

Third, model explainability is essential for building trust in AI-driven medical diagnostics. The integration of techniques such as Gradient-weighted Class Activation Mapping (Grad-CAM) or SHapley Additive exPlanations (SHAP) would enhance the interpretability of model predictions, allowing clinicians to better understand the decision-making process. By providing visual explanations for predictions, these methods can improve transparency and facilitate clinical adoption. Finally, regulatory compliance is a key consideration for real-world deployment. AI-based medical tools must undergo rigorous validation to align with FDA and CE regulatory standards, ensuring their safety, efficacy, and reliability in clinical settings. Future work should focus on bridging the gap between AI model development and regulatory approval to facilitate seamless integration into healthcare systems.

This augmentation process can help the model generalize better and handle unseen samples by exposing it to a wider range of scenarios. Thirdly, running the same datasets and using similar parameters in other similar platforms like Google's Vertex AI would allow for a comparison of classification results and help identify which application yields better performance. This comparative analysis can provide valuable insights into the strengths and weaknesses of different platforms and inform the selection of the most effective tool for COVID-19 prognosis classification. Additionally, incorporating clinical data or additional diagnostic information, such as patient demographics or laboratory results, could further enhance the model's predictive capabilities. By integrating these additional inputs, the model can learn to leverage relevant contextual information and improve its accuracy in predicting COVID-19 prognosis based on X-ray images.

Finally, continuous model evaluation and monitoring, accompanied by regular updates and retraining, are essential to ensure the model's adaptability to evolving data distributions and emerging patterns. By continuously assessing the model's performance and incorporating new data, the model can maintain its effectiveness over time and adapt to changes in COVID-19 manifestations and imaging characteristics. Addressing these potential areas of improvement can lead to enhanced performance and accuracy in COVID-19 prognosis classification, ultimately benefiting clinical decision-making and patient care.

While the integration of deep learning models in clinical settings for COVID-19 prognosis classification offers significant benefits, it is essential to consider the ethical concerns associated with the use of AI models in patient diagnosis and healthcare management. One key concern revolves around patient privacy and the secure handling of sensitive medical data [55]. Deep learning models rely on vast amounts of patient information, including X-ray images and medical records, to make accurate predictions. Ensuring the privacy and confidentiality of patient data becomes paramount to maintain trust and comply with ethical guidelines [56]. Robust data anonymization techniques, strict access controls, and adherence to data protection regulations are crucial to safeguard patient privacy.

Moreover, the deployment of AI models raises questions about the transparency and interpretability of their decisions. Deep learning models often operate as black boxes, making it challenging to understand the underlying factors influencing their predictions. In a clinical context, where decisions can have significant consequences for patients, the ability to explain and justify the rationale behind AI-driven diagnoses becomes crucial. Efforts are being made to develop explainable AI techniques that provide insights into the decision-making process of these models, enabling clinicians to trust and validate their outputs. Another ethical concern relates to the potential biases present in the training data used to develop deep learning models. If the training data is skewed or fails to represent the diversity of patient populations, the model's performance may be compromised, leading to disparities in diagnostic accuracy across different demographics. It is essential to address and mitigate these biases to ensure equitable and fair healthcare outcomes for all patients.

In summary, while deep learning models hold great promise in clinical diagnosis and patient management, careful attention must be given to the ethical implications surrounding privacy, transparency, and bias. By addressing these concerns through robust privacy protocols, explainable AI techniques, and bias mitigation strategies, healthcare professionals can harness the full potential of AI models while upholding the highest ethical standards in patient care.

**CONCLUSION**

Our findings confirm that deep learning models trained on Microsoft Azure Custom Vision can effectively classify COVID-19 prognosis based on X-ray images, with statistical validation and effect size analyses demonstrating that additional training significantly enhances model performance, particularly for balanced datasets. Set 3 - Iteration 2 achieved the highest accuracy (97%), reinforcing the impact of dataset augmentation and extended training. However, performance variations across different iterations highlight the need for robust dataset curation and external validation to ensure model generalizability.

The application of deep learning models for COVID-19 prognosis classification has substantial implications for clinical practice. By automating the analysis of X-ray images, these models can reduce the burden of manual image interpretation, leading to faster diagnosis, improved triaging, and more efficient treatment planning. The ability to accurately classify COVID-19 cases into mild, moderate, and severe categories provides clinicians with critical insights that can assist in prioritizing patient management and optimizing healthcare resources.

Despite these promising results, challenges remain in accurately classifying certain categories, particularly mild and moderate cases. This underscores the need for further refinements in model architecture, dataset expansion, and explainability techniques to improve performance and clinical applicability. Future work should focus on validating the model with external datasets, benchmarking against state-of-the-art deep learning architectures (e.g., ResNet, EfficientNet, Vision Transformers), and integrating interpretability tools such as Grad-CAM and SHAP analysis. Additionally, efforts should be made to align AI-based diagnostic tools with regulatory standards (e.g., FDA, CE) to facilitate clinical adoption.

Overall, our study demonstrates that deep learning models can serve as valuable tools for COVID-19 prognosis classification, supporting clinicians in decision-making, risk assessment, and treatment optimization. By addressing current limitations and incorporating advancements in model generalization, interpretability, and regulatory compliance, AI-driven diagnostic solutions have the potential to play an integral role in enhancing healthcare outcomes and pandemic preparedness.

**DATA AVAILABILITY**

The X-ray datasets for this project are openly available on https://github.com/alfred5063/xraydatasets, providing accessibility for further research and development. These datasets facilitate advancements in medical imaging and AI-driven diagnostics. Additionally, the trained model can be accessed and tested at https://covintec.streamlit.app/, offering an interactive platform for evaluating its performance. This ensures transparency, reproducibility, and practical application in real-world scenarios.